\documentclass{iucr}              
\RequirePackage{graphicx}
\usepackage{amsmath}
\usepackage{amssymb}
\usepackage{upgreek}
\usepackage{color}
     \journalcode{J}              

\begin{document}                  



\title{A computationally efficient method to solve the Takagi-Taupin equations for a large deformed crystal}


\cauthor[a]{Ari-Pekka}{Honkanen}{ari-pekka.honkanen@helsinki.fi}{}
\author[b]{Giulio}{Monaco}
\author[a]{Simo}{Huotari}

\aff[a]{Department of Physics, P.~O.~Box 64, FI-00014 Helsinki, \country{Finland}}
\aff[b]{Physics Department, University of Trento, Via Sommarive 14, 38123 Povo (TN), \country{Italy}}









\maketitle                        

\begin{synopsis}
\end{synopsis}

\begin{abstract}
We present a treatise on solving the Takagi-Taupin equations in the case of a strain field with an additional, spatially slowly varying component (owing to \emph{e.g.}~heat expansion or angular compression). We show that
the presence of such a component in a typical case merely shifts the reflectivity curve as a function of wavelength or incidence angle, while having a negligible effect on its shape. On the basis of the derived result, we develop a computationally efficient method to calculate the reflectivity curve of a large deformed crystal. 
The validity of the method is demonstrated by comparing computed reflectivity curves with experimental ones for bent silicon wafers. A good agreement is observed.
\end{abstract}


\section{Introduction}

In the hard X-ray regime, the X-ray spectrometers with the highest energy resolution  are nowadays based on diffractive crystal optics.
In the sub-eV energy resolution range, bent crystals are often used to yield an optimal collection solid angle and a suitable bandwidth. There are numerous different curved crystal geometries in common use, such as the Johann, Johansson, and von Hamos geometries \cite{johann,johansson,vonhamos}.
X-ray crystal spectrometers based on such designs are installed at various synchrotron light sources worldwide \emph{e.g.}~\cite{itou01,fister06,verbeni09,hiraoka13,sokaras13,alonsomori15,rueff15}.
 

To design the most precise instruments, one requires a solid theoretical knowledge of the diffraction properties of crystals. To this end, highly relevant for bent crystals was the theory of dynamical x-ray diffraction in deformed crystals that was developed
 independently by S. Takagi and D. Taupin \cite{takagi1,takagi2,taupin}. 
In the heart of the theory are so called Takagi-Taupin (TT) equations that describe the wave field in a (quasi)periodic medium. For the usual case of two-beam diffraction, one obtains the X-ray reflectivity curve of a crystal by solving a pair of partial differential equations. When the strain field is solely depth-dependent, these equations reduce to a one ordinary differential equation. This is the usual approach to model the X-ray reflectivity of bent crystals, but it is not enough e.g. for spherically bent crystals used in Johann type spectrometers owing to a spatially slowly changing component of strain arising from so called angular compression \cite{verbeni09,honkanen14,honkanen14_2}. Accurate numerical methods have been developed \cite{authier68,yan14},  yet solving the general two-beam TT-equations for a crystal wafer of typical size of $\sim$100~mm is a computational challenge. Another important
example of the situations that require solving the TT-equations over a large crystal area is X-ray monochromators with heat load induced deformations \cite{hoszowska01,zhang13}.

In this paper, we examine the solutions of the TT-equations and as a result present an efficient method for computing the reflectivity curve of a large deformed crystal. The method applies to strain fields that can be decomposed to a sum of a depth-dependent component that varies rapidly along the path of the incident and the diffracted beams and a slowly changing component that varies in macroscopic scale. The result generalises our earlier procedure presented in \cite{honkanen14}.

\section{TT-equations with a slowly changing strain component}

The Takagi-Taupin equations that describe two-beam diffraction in a crystal are \cite{gronkowski91}
\begin{equation}
\begin{cases}
 \cfrac{\partial D_0}{\partial s_0} = -\pi i k \chi_0 D_0 - \pi i k C \chi_{\bar{h}} D_h \\
 \cfrac{\partial D_h}{\partial s_h} = -\pi i k C \chi_h D_0 + 2 \pi i k \beta_h D_h
\end{cases},\label{eq:TT}
\end{equation}
where $D_0$ and $D_h$ are the amplitudes of the incident and diffracted waves inside the crystal, 
$\partial/\partial s_0$ and $\partial/\partial s_h$ are derivatives with respect to the directions of the forward diffracted
and diffracted waves, $k$ is the wave number of the incident wave, $\chi_0$ and $\chi_h$ are the Fourier components of the susceptibility
corresponding to reciprocal lattice vectors $\mathbf{0}$ and $\mathbf{h}$, and $C$ is the polarisation factor.
$\beta_h$ is given by
\begin{align}
\beta_h = \frac{1}{2} \left(\frac{|\mathbf{k}+\mathbf{h}-\nabla ( \mathbf{u}\cdot\mathbf{h})|^2-k^2}{k^2} - \chi_0\right)\\
= \frac{|\mathbf{h}|^2}{2 k^2} + \frac{\mathbf{k}\cdot\mathbf{h}}{k^2} - \frac{\chi_0}{2} - \frac{1}{k}\frac{\partial (\mathbf{u}\cdot\mathbf{h})}{\partial s_h},\label{eq:beta_init}
\end{align}
where $\mathbf{k}$ is the wave vector of the incident wave and $\mathbf{u}$ is the displacement field. The reflectivity curve of an arbitrarily deformed crystal is obtained by varying either direction or length of $\mathbf{k}$ and solving the TT-equations in the vicinity of the Bragg condition.

For solving the TT-equations, $\beta_h$ is typically reformulated in terms of the Bragg angle and the angular deviation from the Bragg condition. However, as the Bragg angle is not defined for wavelengths smaller than the backscattering wavelength, this formulation ceases to be valid in near-backscattering conditions \cite{caticha82}. 
In this paper, we circumvent the arising problems by formulating \eqref{eq:beta_init} in terms of the wavelength $\lambda$ and the incidence angle $\theta$ so that
\begin{equation}
\beta_h
= \frac{\lambda^2}{2 d_h^2} - \frac{\lambda}{d_h} \sin \theta - \frac{\chi_0}{2} - \lambda \frac{\partial (\mathbf{u}\cdot\mathbf{h})}{\partial s_h},\label{eq:beta}
\end{equation}
where $d_h$ stands for the separation of the diffractive Bragg planes.

As one can see, the effect of the strain field comes into the TT-equations via parameter $\beta_h$. Thus, the radiation propagation is identical in crystals with different strain fields if $\beta_h$ are exactly alike. One can tune the value of $\beta_h$ by changing 
the angle of incidence $\theta$ or the wavelength $\lambda$ of the radiation. 
This is approximately true even in the latter case even if a change of $\lambda$ affects the solutions of Eqs.~\eqref{eq:TT} in a non-trivial way, since the relative shift of $\lambda$ owing to the strain is practically negligible.


Let us consider the diffraction of a crystal in two different cases. In the first one the displacement field is $\mathbf{u}_{I }= \mathbf{u}_f(z)$, which accounts for a rapid displacement for the depth-dependent strain. 
In the other case the displacement field is 
$\mathbf{u}_{II}=\mathbf{u}_f + \mathbf{u}_s$, where the 
component of the strain described by $\mathbf{u}_s=\mathbf{u}_s(x,y)$ is nearly linear in terms of $s_h$ in microscopic scale, thus representing the slowly varying component of the strain. The coordinate system is chosen so that the $z$-direction is normal to the crystal surface, positive direction being outward from the crystal.
The strain field of a spherically bent crystal analyser serves as an example of the latter kind of displacement field \cite{honkanen14}.  
Subscripts $I$ and $II$ will be used throughout 
the article to refer to the quantities relating to the diffraction of 
these cases, respectively. Substituting $\mathbf{u}_{\mathrm{I}}$ and $\mathbf{u}_{\mathrm{II}}$ into \eqref{eq:beta}
and taking their difference, we get
\begin{align}\label{eq:simcond}
 \Delta \beta_h &= \frac{\lambda_{II}^2-\lambda_{I}^2}{2 d_h^2}
 -\left( \frac{\lambda_{II}}{d_h} \sin \theta_{II} -
 \frac{\lambda_{I}}{d_h} \sin \theta_{I}\right)
   \nonumber \\ 
&-\lambda_{II} \frac{\partial(\mathbf{u}_s\cdot\mathbf{h})}{\partial s_h}-\left(\lambda_{II}-\lambda_{I}\right) \frac{\partial(\mathbf{u}_f\cdot\mathbf{h})}{\partial s_h}.
\end{align}
On the basis of the earlier argument, the solution to the TT-equation 
is found to be equivalent in both cases if the similarity condition 
$\Delta \beta_h = 0$ holds.
In the following, the condition is met via a constant change in the incidence angle or wavelength under certain assumptions. The two cases are examined separately and the results are applied to compute the reflectivity curves.

\subsection{Scanning the incident wavelength}
Let us first examine the case of a varying wavelength. We denote the difference of wavelengths between cases $I$ and $II$ with $\delta \lambda \equiv \lambda_{II}-\lambda_{I}$. In this case the angle of incidence is kept constant \emph{i.e.}~$\theta_I=\theta_{II}\equiv\theta$.

\begin{figure}\label{fig:vectors}
\caption{Relations of $\mathbf{s}_0$, $\mathbf{s}_h$, $\mathbf{s}_{\parallel}$, $\mathbf{s}_{\perp}$, and $\mathbf{h}$ to each other.}
\scalebox{0.8}{\includegraphics{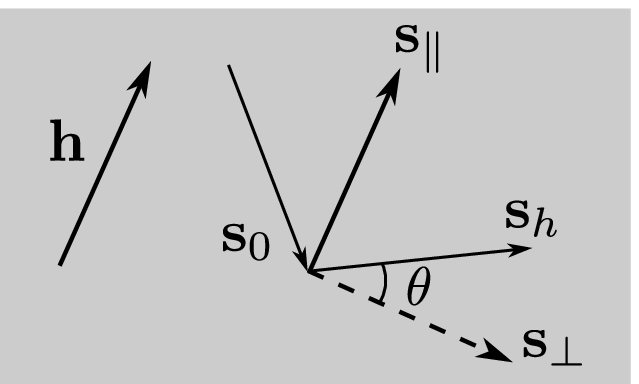}}
\end{figure}

Since the differences caused by $\mathbf{u}_s$ are small, we can make the first-order approximation
\begin{equation}
\lambda_{II}^2 \approx \lambda_{I}^2 + 2 \lambda_{I} \delta \lambda.
\end{equation}
Thus the similarity condition can be written as
\begin{equation}\label{eq:cond1}
\frac{\lambda_I \delta \lambda}{d_h^2}-\frac{\delta \lambda}{d_h}\sin \theta
-\lambda_I \frac{\partial(\mathbf{u}_s\cdot\mathbf{h})}{\partial s_h}
-\delta \lambda \frac{\partial(\mathbf{u}_{II}\cdot\mathbf{h})}{\partial s_h} =0.
\end{equation}
We define a new basis with vectors $\mathbf{s}_{\parallel}$ and $\mathbf{s}_{\perp}$ being respectively parallel and perpendicular to $\mathbf{h}$ as shown in Figure~\ref{fig:vectors}.
The first partial derivative in \eqref{eq:cond1} can be now written as
\begin{equation}\label{eq:epstau}
\frac{\partial(\mathbf{u}_s\cdot\mathbf{h})}{\partial s_h}= \frac{\epsilon \sin \theta +\tau \cos \theta}{d_h},
\end{equation}
where $\epsilon \equiv\partial(\mathbf{u}_s\cdot\mathbf{\hat{h}})/\partial s_{\parallel}$ is 
the normal strain in direction of $\mathbf{h}$ and $\tau \equiv\partial(\mathbf{u}_s\cdot\mathbf{\hat{h}})/\partial s_{\perp}$ is 
the shear strain in $\mathbf{s}_{\parallel}$-$\mathbf{s}_{\perp}$-plane.
Substituting the former to \eqref{eq:cond1} and neglecting $\delta \lambda \partial(\mathbf{u}_{II}\cdot\mathbf{h})/\partial s_h$ as second-order, we get
\begin{equation}\label{eq:cond2}
\left(\frac{1}{d_h}- \frac{\sin \theta}{\lambda_{I}}\right) \delta \lambda-\epsilon \sin \theta-\tau \cos \theta=0.
\end{equation}
Since the diffraction takes place in the vicinity of the Bragg condition, we can use Bragg's law to express $d_h$ with high accuracy in terms of $\lambda_I$ and $\theta_I$:
\begin{equation}
d_h = \frac{\lambda_I}{2 \sin \theta}.
\end{equation}
Substituting the former to \eqref{eq:cond2}, the similarity condition becomes
\begin{equation}\label{eq:simcond_lambda}
 \frac{\delta \lambda}{\lambda_I} = \epsilon + \tau \cot \theta.
\end{equation}
Notably in the case of exact backscattering or when $\tau = 0$ this reduces to the result expected from Bragg's law that was used in \cite{honkanen14}.

From Equation~\eqref{eq:simcond_lambda} we now see that the change of the wavelength shift $\Delta (\delta \lambda)$ relative to the width of the diffraction peak $\Delta \lambda_I$ is
\begin{equation}\label{eq:simcond_deltalambda}
 \frac{\Delta(\delta \lambda)}{\Delta \lambda_I} = \epsilon + \tau \cot \theta,
\end{equation}
which is typically $\lesssim 10^{-4}$ supposing we are not close to grazing-incidence conditions. This means that effect of additional strain components $\epsilon$ and $\tau$ present in case $II$ are accountable by simply shifting the reflectivity curve of case $I$ in wavelength domain by an amount dictated by Equation~\eqref{eq:simcond_lambda}.

The validity of the derivation was studied using a \emph{Python/SciPy} implemented one-dimensional TT-solver.\footnote{The solver is freely available under MIT license at \textsc{https://github.com/aripekka/pytakagitaupin}} The derivation of 
the depth-dependent equation was according to \cite{gronkowski91} with the exception of using \eqref{eq:beta} as the form of $\beta_h$ in the derivation. The reflectivity curve for symmetric (660)-reflection of silicon was computed at incidence angles of 85$^\circ$, 75$^\circ$, 65$^\circ$ and 55$^\circ$ (approx. photon energies of 9.72~keV, 10.0~keV, 10.7~keV, and 11.8~keV, respectively) 
for a set of constant strains $\epsilon$ varying from 0 to 10$^{-3}$. $\tau$ was set to zero. The thickness of the crystal was set to 3~mm in order to get rid of the thickness related oscillations in the reflectivity curve which interfere with the accurate determination of the curve width.
The shift of the reflectivity maximum and full widths at half maxima (FWHMs) as function of $\epsilon$ were computed and the results were compared with the theoretical predictions of Equations \eqref{eq:simcond_lambda} and \eqref{eq:simcond_deltalambda}.

\begin{figure}\label{fig:lambdashifts}
\caption{Simulated and theoretical shifts in the wavelength as a function of strain component $\epsilon$ at different incidence angles for symmetrical Si(660)-reflection. Some of the data points are not visible owing to their overlap.}
\scalebox{0.45}{\includegraphics{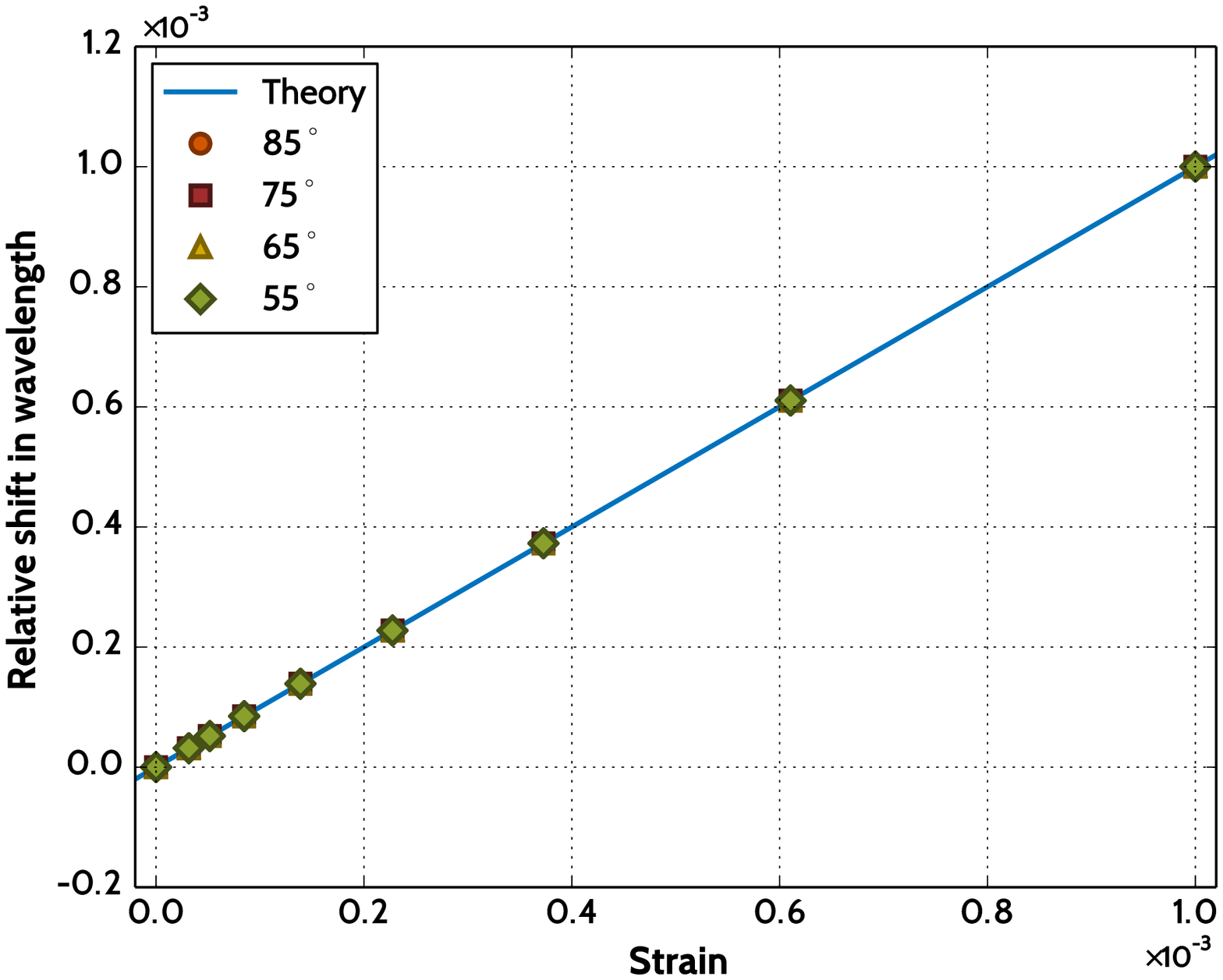}}
\end{figure}

\begin{figure}\label{fig:lambdafwhmshifts}
\caption{Simulated and theoretical changes in the FWHM of the reflectivity curve as a function of strain component $\epsilon$ at different incidence angles for symmetrical Si(660)-reflection. Some of the data points are not visible owing to their overlap.}
\scalebox{0.45}{\includegraphics{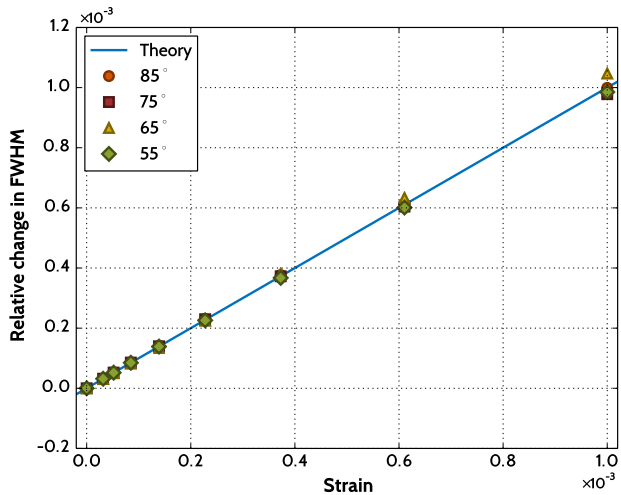}}
\end{figure}

The relative shifts in the wavelength as a function of strain $\epsilon$ 
is presented in Figure~\ref{fig:lambdashifts} with the theoretical prediction. As one can see, the predicted shift is found to be in  good accordance with the simulations, and follows the behavior expected simply from Bragg's law.
The shape and the width of the reflectivity curve have much weaker dependence on the strain, as shown in Figure~\ref{fig:lambdafwhmshifts} for the FWHM. As in the case of the wavelength shift, the simulated results are found to follow the theoretical result. 

\subsection{Scanning the incidence angle}
Another approach is offered by a variation of the beam's angle of incidence with respect to the crystal surface. We denote the difference of incidence angles 
between cases $I$ and $II$ with $\delta \theta \equiv \theta_{II}-\theta_{I}$. In this case 
the wavelength is kept constant \emph{i.e.}~$\lambda_I=\lambda_{II}\equiv\lambda$. 
From Equation~\eqref{eq:simcond}, the similarity condition now becomes
\begin{equation}
\frac{1}{d_h}\left(\sin \theta_{II}-\sin \theta_{I}\right) +  \frac{\partial(\mathbf{u}_s\cdot\mathbf{h})}{\partial s_h} = 0.
\end{equation}
Acknowledging that angles in Eq.~\eqref{eq:epstau} correspond to the case $II$, we obtain by substitution
\begin{equation}\label{eq:theta_cond}
\sin \theta_{II}-\sin \theta_{I} + \epsilon \sin \theta_{II}
+\tau \cos \theta_{II} = 0.
\end{equation}
By making the Taylor expansion in terms of $\delta \theta$ and retaining only the first-order terms we obtain for the similarity condition
\begin{equation}\label{eq:simcond_theta}
\delta \theta = - \epsilon \tan \theta_I - \tau.
\end{equation}
The change in the incidence angle shift $\Delta(\delta \theta)$ relative to the angular width of the diffraction $\Delta \theta_I$ is found by differentiation of Eq.~\eqref{eq:simcond_theta}:
\begin{equation}\label{eq:simcond_deltatheta}
\frac{\Delta(\delta \theta)}{\Delta \theta_I} = -\frac{\epsilon}{\cos^2 \theta_I}.
\end{equation}
As seen in the case of wavelength, the shift in the angle of incidence is also found to be linear in terms of the strain 
components $\epsilon$ and $\tau$. 
However, in this case the first-order expressions diverge when $\theta=90^\circ$.
In addition, since \eqref{eq:simcond_deltatheta} grows faster than \eqref{eq:simcond_theta} when $\theta$ approaches 90$^\circ$, 
the change in the width of the rocking curve can not be necessarily neglected. In the vicinity of backscattering, one should consider a 
higher-order expansion of Equation~\eqref{eq:theta_cond} in terms of $\delta \theta$. This is, however, out of the scope of this work as
the situation becomes more complicated owing to the symmetricity of the reflection at angles above 90$^\circ$.

Validity of the derivation was examined using the one-dimensional TT-solver as for the wavelength. Si(660)-reflection was studied at the photon
energies of 9.72~keV, 10.0~keV, 10.7~keV, and 11.8~keV, corresponding approximately to the incidence angles of 85$^\circ$, 75$^\circ$, 65$^\circ$ and 55$^\circ$, respectively. 
The constant strain component $\epsilon$ varied from 0 to 10$^{-3}$ and $\tau$ was set to zero. 

\begin{figure}\label{fig:thetashifts}
\caption{Simulated and theoretical shifts in the incidence angle as a function of strain component $\epsilon$ for different photon energies for symmetrical Si(660)-reflection. The data is divided by $-\tan \theta$ for clarity. Some of the data points are not visible owing to their overlap.}
\scalebox{0.45}{\includegraphics{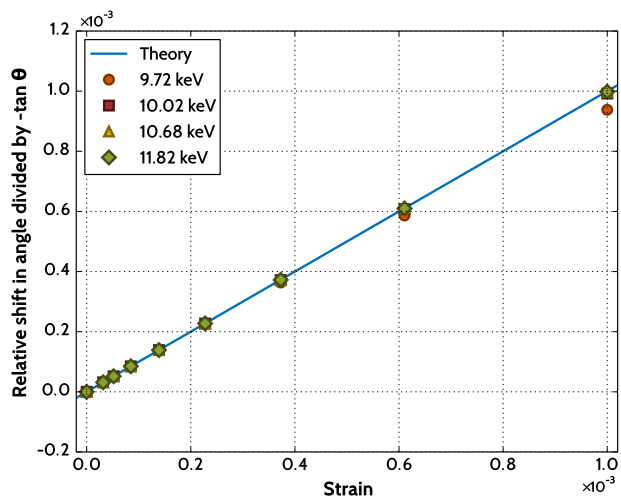}}
\end{figure}

\begin{figure}\label{fig:thetafwhmshifts}
\caption{Simulated and theoretical changes in the FWHM of the rocking as a function of strain component $\epsilon$ for different photon energies for symmetrical Si(660)-reflection. The data is multiplied  by $-\cos^2 \theta$ for clarity. Some of the data points are not visible owing to their overlap.}
\scalebox{0.45}{\includegraphics{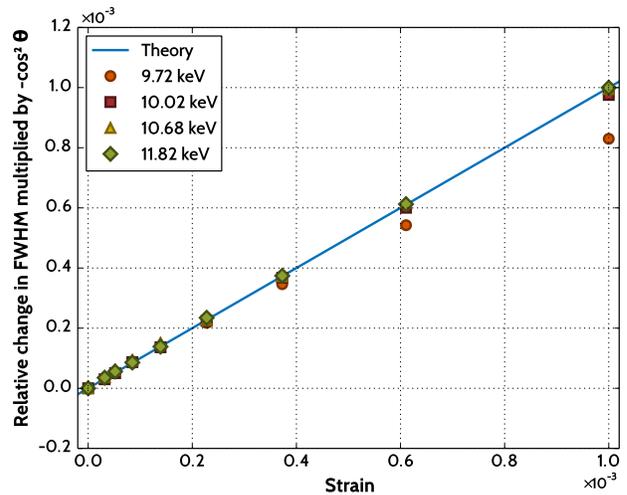}}
\end{figure}

The results are presented in Figures~\ref{fig:thetashifts} and \ref{fig:thetafwhmshifts}. As in the case of wavelength, the simulated 
$\delta \theta$ follow the theoretical prediction \eqref{eq:simcond_theta} with high precision and the general shape of the curve appears 
to be independent of $\epsilon$. Also FWHMs of the rocking curves are in good 
accordance with Equation~\eqref{eq:simcond_theta}. 
At a higher Bragg angle at 9.72~keV, the simulated values begin deviate from the theory as $\epsilon \tan \theta_I$ and $\epsilon \cos^2 \theta_I$ becomes larger. Apart from the non-linear regime, the theory holds well for the smaller values of $\epsilon$.

\section{Efficient computation of the reflectivity curves}

Based on the results on the previous section, the effect of a slowly changing strain field is locally taken into account by 
a simple shift of the solution to TT-equations on either the wavelength or incidence angle scales. The width of the curve is also altered slightly but as the relative change is expected to be $< 1$\%, it can be neglected in most cases. The reflectivity curve of the whole crystal is 
then obtained by summing up the reflectivities of infinitesimal areas over the crystal surface. Since solving the TT-equations
can be computationally demanding for a macroscopic crystal, the derived results offer an
intriguing method of computation in cases where a suitable strain component is present.

The reflectivity curve of a macroscopic crystal can be solved as follows. 
The TT-equations \eqref{eq:TT} are solved in  their one-dimensional form for the depth-dependent displacement component $\mathbf{u}_f$ as a function of $\theta$ or $\lambda$.
The slowly changing component $\mathbf{u}_s$ is used to compute the $\delta \lambda$- 
or $\delta \theta$-distribution over the crystal surface using Eq.~\eqref{eq:simcond_lambda} or Eq.~\eqref{eq:simcond_theta}, respectively. 
The reflectivity curve of the crystal is then obtained by convolving 
the TT-curve with $\delta \lambda$- or $\delta \theta$-distribution. Other 
contributions, such as bandwidth of the X-rays or geometric factors, are convolved
with the result as needed.

\section{Experimental verification}

\begin{figure}\label{fig:analyser}
\caption{Geometry of the studied Si(660) analysers. 300 $\mu$m thick silicon wafers with the diameter of 100 mm were anodically bonded on a spherical glass concave. The bending radius was 1~m. Two cuts were made symmetrically along the [001]-direction limiting the ana\-lyser dimension in the [1$\bar{1}$0]-direction to 80~mm. }
\scalebox{.7}{\includegraphics{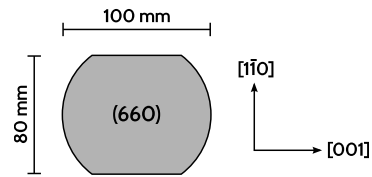}}
\end{figure}

We applied the method to compute the reflectivity curve 
of the anodically bonded spherically bent Si(660) analyser of a specific kind of geometry 
presented in Figure~\ref{fig:analyser} at the incidence angle of 88.7$^{\circ}$. The bending 
radius of the crystals was 1~m and the thickness of the wafers was 300~$\upmu$m. The strain field of 
the wafer is calculated according to \cite{honkanen14}.
The $\delta \lambda$-distribution was computed using Equation~\eqref{eq:simcond_lambda} omitting the term containing $\tau$ as we are close to back-scattering. The depth-dependent TT-curve was convolved with $\delta \lambda$-distribution and a gaussian with FWHM of 235 meV to take into account the bandwidth of the incident radiation in the experimental setup.
We compared the theoretical prediction to the experimental 
curves of two of such analysers measured at the
inelastic X-ray scattering beamline ID20 at the ESRF (European Synchrotron Radiation Facility). The details of the experimental setup are presented in \cite{honkanen14_2}. 

\begin{figure}\label{fig:results}
\caption{The measures reflectivity curves of two cut Si(660) analysers (solid lines) compared with the theoretical prediction (black dashed line). The red dashed line shows the predicted reflectivity in the case where the slowly changing strain component is not taken into account.}
\scalebox{0.45}{\includegraphics{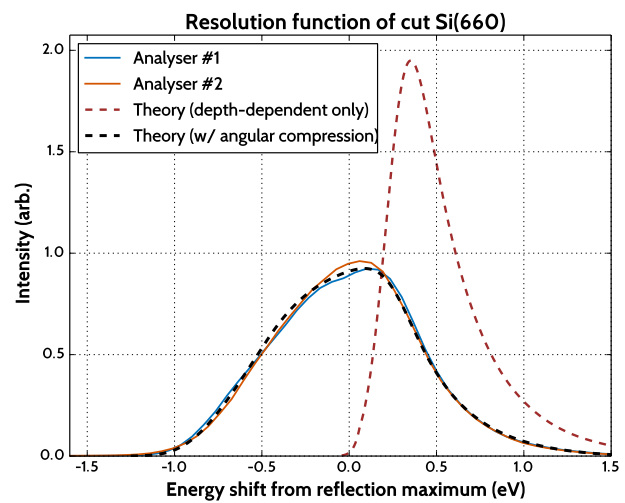}}
\end{figure}

The comparison between the theory and the experiment is presented in Figure \ref{fig:results}. The intensities of the curves are normalised with respect to the integrated intensity.
The horizontal axis is the energy difference between the incident-photon energy and centroid of the measured reflectivity curve.\footnote{
The absolute energy can in principle be determined as well, but as we concentrate on the use of the method in energy-loss spectroscopies, the relative changes on the energy shift axis are more relevant.} As it is seen, a good agreement 
is found between the theory and the experiment.For comparison, the red dashed curve shows the predicted reflectivity when the slowly changing strain component owing to angular compression is neglected. It is evident that the slowly changing component changes the shape of the curve to such extent it can not be simply overlooked in the case of a macroscopic analyser crystal.It has a tendency to increase the spectral weight at the energy gain side (negative energy shift) of the spectrum, and hence creates an apparent shift of the reflectivity curve in this example by $\sim0.5$~eV.

\section{Conclusions}

In this paper we have examined how solutions of TT-equations behave in presence of a slowly varying component of the strain field. We applied the results to construct an efficient semi-analytical method to compute the X-ray reflectivity of an deformed crystal with a slowly varying strain component. 

We used the method to compute the reflectivity curve of a Si(660) analyser cut in a specific way. Compared to the measured reflectivity curves, a precise correspondence is found. Such an agreement found in the case of the examined non-trivial geometry speaks for the predictive power of the presented method which offers an appealing alternative for effective computation of the reflectivity curves of large, deformed crystals.




\ack{Acknowledgements}
The authors would like to thank Roberto Verbeni, Laura Simonelli, Marco Moretti Sala, Ali Al-Zein and Michael Krisch 
for the experimental data of the bent silicon analysers. A-PH and SH were funded by the Academy of Finland (grants 1254065, 1283136 and 1259526).

\bibliographystyle{iucr}
\bibliography{article}





\end{document}